\documentclass[letterpaper, 10 pt, conference]{ieeeconf}

\IEEEoverridecommandlockouts
\usepackage{cite}

\usepackage[pagewise]{lineno}
\usepackage{graphicx}
\graphicspath{{./figs/}}

\usepackage{multirow}

\usepackage{caption} %added so the algo I included doesn't say fig..
\usepackage{graphics} % for resizing table
\usepackage{adjustbox}
\usepackage{comment}
\usepackage[utf8]{inputenc}
\usepackage[T1]{fontenc}
\usepackage[cmex10]{amsmath}
\usepackage{algorithm}
\usepackage{algpseudocode}
\usepackage{amssymb}

\mathchardef\mhyphen="2D

\usepackage{subfigure}

\title{\LARGE \bf
Collision Detection for Multi-Robot Motion Planning\\
with Efficient Quad-Tree Update and Skipping
}
\author{Abdel Zaro$^1$, Ardalan Tajbakhsh$^2$ and Aaron M. Johnson$^2$% <-this % stops a space
\thanks{*This work was supported in part by the National Science Foundation under Grant IIS-1659774 and the RISS program.}% <-this % stops a space
    \thanks{$^1$ Department of Mechanical Engineering, UC Berkeley $^2$ Department of Mechanical Engineering, Carnegie Mellon University, Pittsburgh, PA, USA, \texttt{(atajbakh, amj1)@andrew.cmu.edu}}%
}
\begin{document}
%\linenumbers % Uncomment this to enable line numbers in the peer review
\bstctlcite{IEEEexample:BSTcontrol} %This line, along with the text at the top of the .bib file, shortens citations to use et al

\maketitle
\thispagestyle{empty}
\pagestyle{empty}

%%%%%%%%%%%%%%%%%%%%%%%%%%%%%%%%%%%%%%%%%%%%%%%%%%%%%%%%%%%%%%%%%%%%%%%%%%%%%%%%
\begin{abstract}
This paper presents a novel and efficient collision checking approach called \textit{Updating and Collision Check Skipping Quad-tree (USQ)} for multi-robot motion planning. USQ extends the standard quad-tree data structure through a time-efficient update mechanism, which significantly reduces the total number of collision checks and the collision checking time. In addition, it handles transitions at the quad-tree quadrant boundaries based on worst-case trajectories of agents. These extensions make quad-trees suitable for efficient collision checking in multi-robot motion planning of large robot teams. We evaluate the efficiency of USQ in comparison with Regenerating Quad-tree (RQ) from scratch at each timestep and naive pairwise collision checking across a variety of randomized environments. The results indicate that USQ significantly reduces the number of collision checks and the collision checking time compared to other baselines for different numbers of robots and map sizes. In a 50-robot experiment, USQ accurately detected all collisions, outperforming RQ which has longer run-times and/or misses up to 25\% of collisions.

% In particular, for a 50 robot experiment USQ reduces the collision checking time by approximately 25\%, 50\%, and 145\% compared to RQ, NSQ, and pairwise collision checking. 
\end{abstract}

\begin{keywords}
Multi-Robot Systems, Motion Planning for Multiple Mobile Robots, Collision Detection
\end{keywords}  

%%%%%%%%%%%%%%%%%%%%%%%%%%%%%%%%%%%%%%%%%%%%%%%%%%%%%%%%%%%%%%%%%%%%%%%%%%%%%%%%
\section{INTRODUCTION}

\begin{comment}
\begin{figure}[!t]
    \centering 
    \includegraphics[width=.8\linewidth]{figs/fig1.png}
    \caption{On the left, both updating the quad-tree and collision checking within the node is skipped for one timestep at $t=2$. Skipping occurs when all the robots are both far from the boundary by staying out of the blue dashed line area and far from each other. At $t=3$, skipping does not occur because robot A is near the boundary, illustrated by it being outside the dashed line. On the right, the Regenerating quad-tree (RQ) must update robots in the quad-tree and perform collision checks at each timestep.}
    \label{fig:robot_plans}
\end{figure}
\end{comment}

As robotic systems scale up to many agents operating in a shared environment, path and motion planning become major challenges in practical deployments due to the increasing number of states and constraints from other robots and the environment. In particular, collision checking becomes a major bottleneck in planning as the number of agents increases, since in the worst-case every agent usually has to check collisions with every other agent at every timestep. In receding horizon planning, this becomes even more challenging as collision checking needs to occur over the entire planning horizon at each update. 

The majority of current multi-robot planning methods use pairwise collision checking, which involves checking collisions between every pair of robots at every timestep, as in the conflict-based search algorithm \cite{sharon2012}. This can become quite expensive as the number of robots increases since it scales quadratically with the number of robots \cite{pairwise_big_o_reference}. In addition, many of these collision checks are practically unnecessary, since some agents will never collide in the near future, even in the worst case scenario. 

\begin{figure}[tb]
    \centering 
    \includegraphics[scale=0.5]{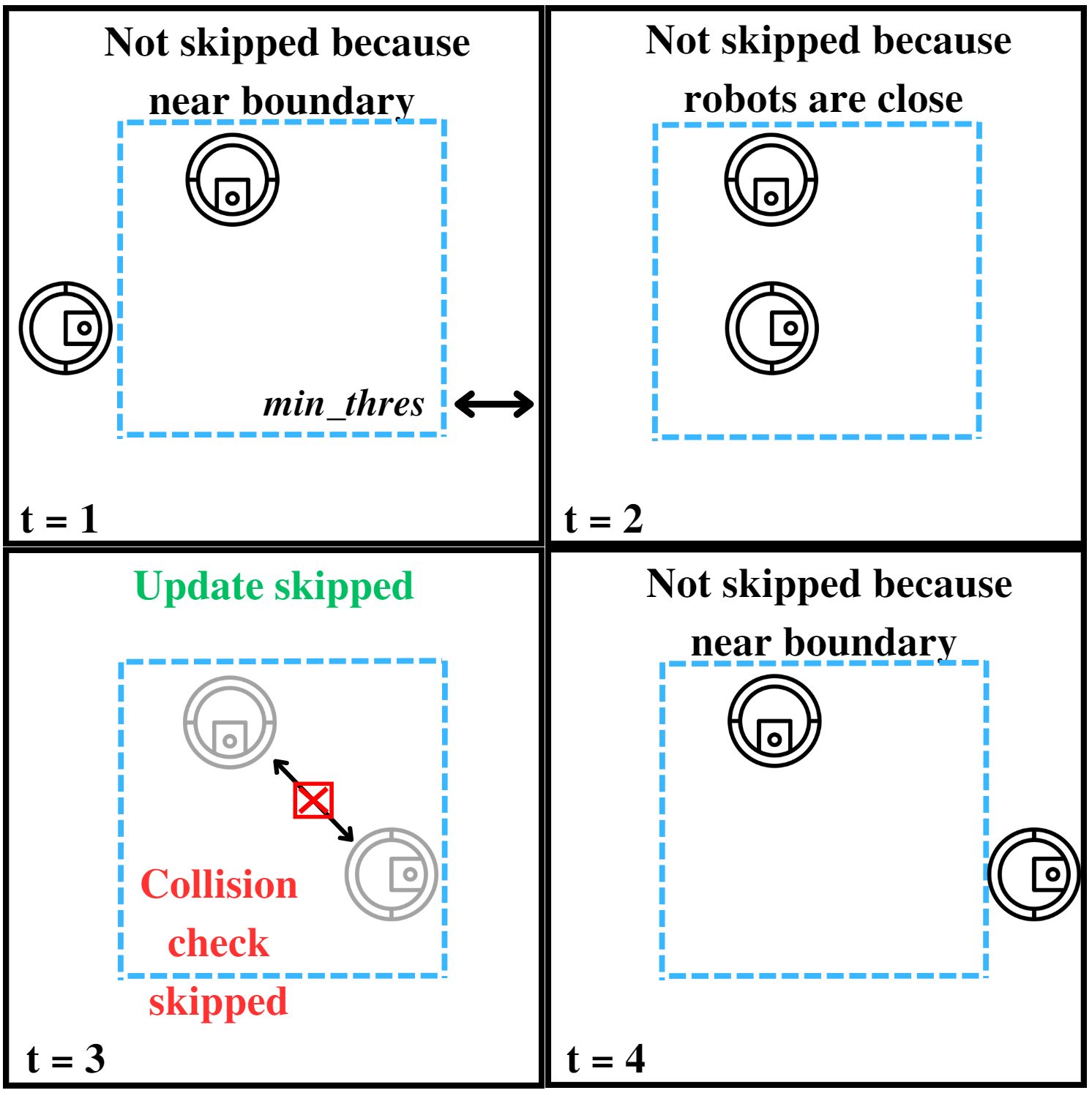}
    \caption{Illustration of the conditions required for update and collision check skipping to take place. In the bottom left quadrant, collision check and update skipping occur because the robots are far from each other and the quadrant's boundary.}
    \label{fig:cases}
 \end{figure}

Other approaches rely on partitioning the free-space efficiently to quickly evaluate whether generated trajectories by the motion planner are collision-free. These methods either approximate the free-space by generating convex polygons over the entire map, or construct robot-centric pyramids over a limited horizon \cite{gao2017gradient,chen2016online,tordesillas2021faster}. While they are effective for fast robot-environment collision checking, they do not handle many robot-robot collisions well. Another class of solutions leverage quad-trees \cite{Botea2004Jan, Kambhampati1986Sep, Kitamura1995Aug}. Quad-trees iteratively partition the map into quadrants until each quadrant holds at most a set number of robots. This allows checking inter-robot collisions for only a subset of robots at any given timestep. 

One major challenge with using trees for applications with multiple moving bodies is efficiently updating the tree in real-time. A few approaches have been proposed to efficiently address this problem (e.g.\ \cite{Ding,Alamri2013Oct}), but they have not been applied to multi-robot collision checking. Furthermore, they require checking tree updates at every timestep. This can become expensive, as each update checking requires traversing the quad-tree to a specific element and looking up its position. Finally, existing methods do not explicitly reason about agents near boundaries and can be prone to missing critical collision checks as agents traverse to neighboring boundaries.

% Octrees, which are quad-trees extended to 3D space have been proven to be an effective collision detection method in path planning [ADD REF]. By systematically partitioning the space into smaller regions, the total number of collision checks is significantly reduced to a subset of robots at any given time. 

We propose a quad-tree update mechanism, Fig.~\ref{fig:cases}, that is time-efficient and is geared towards multi-robot collision checking, called Updating and Collision Check Skipping Quad-tree (USQ). Unlike other methods, our approach allows quad-tree updates to be skipped by leveraging the current state and the predicted trajectories of robots. In addition, USQ skips collisions with robots that are not nearby. We propose additional exception handling rules for robots near boundaries to ensure no critical collision check is skipped. We demonstrate that these extensions significantly improve collision checking time compared to other algorithms without missing any critical collisions and provide a promising approach for collision checking in large robot teams.

This paper is organized as follows. Section \ref{sec:related} summarizes related works, while Section \ref{sec:preliminaries} describes the preliminaries. Section \ref{sec:USQ} presents the detailed design of our USQ algorithm. To showcase the effectiveness of our approach, Section \ref{sec:results} discusses experimental results in various environments. Finally, Section \ref{sec:conclusion} concludes the paper and outlines avenues for future work.

%https://www.overleaf.com/project/63ee70b03567236700a3da07/detacher
\section{RELATED WORKS}
\label{sec:related} 
A Quad-tree is an indexing structure that can efficiently store and retrieve data. Quad-trees have many applications such as collision detection \cite{Honig2018Aug,Wu2022May}, environmental mapping\cite{Dames2012Dec, Collins2021May}, space partitioning \cite{Johansen2003May, Cychowski2005Aug, yijun2021fast} and object tracking \cite{Alamri2013Oct, Singh2017Oct, Datcu1992Nov}.
In the context of collision detection, one of their key advantages lies in their ability to effectively partition the space, thereby establishing relationships between nearby robots. This partitioning method facilitates a clear understanding of the proximity between robots, which allows for collision checking to be limited to robots that are near one another. The reduced number of collision checks allows them to operate in O$(n\,log(n))$ time \cite{log(n)}. In \cite{Botea2004Jan}, a single-robot planning approach is proposed utilizing a framed quad-tree, which places high resolution cells around the boundaries of quadrants containing obstacles to improve the quality of the generated paths. However, this approach increases the memory consumption. \cite{Kambhampati1986Sep,yijun2021fast} use hierarchical map decomposition using quad-trees, which reduces the number of node expansions in an $A^*$ planner without compromising solution quality. But, this approach does not update the quad-tree  online in case of dynamic obstacles and other environment changes. Similar space-partitioning approaches have been applied to 3D applications. Octrees, which operate similar to quad-trees in 3D, have been used for single-robot navigation in constrained environments \cite{Kitamura1995Aug, Rodenberg2016Sep}.

Quad-tree-based approaches have also been used in multi-robot settings. \cite{Zhang2022} presents a  multi-robot planning algorithm operating in discrete space that utilizes quad-tree map division. Similar to an occupancy grid, nodes are labeled as free or occupied. For collision avoidance, at every timestep, the robots check whether the node they are heading into is occupied. In \cite{Honig2018Aug}, a discrete-time multi-robot planning algorithm that uses octrees to partition the 3D space for planning and collision detection is presented. It precomputes conflicts and creates safety corridors, which are regions in which collision avoidance is guaranteed. \cite{Wu2022May} introduces a flexible octree structure that permits a less rigid relationship between child and parent nodes. This non-rigid structure improves collision detection and path planning by employing a multi-resolution grid. However, the paper focuses on pre-flight planning rather than real-time execution. In \cite{csenbacslar2019robust}, an octree-based 3D occupancy grid is used for multi-robot planning. The robots do not communicate and collision avoidance is achieved using buffered Voronoi cells, which do not require collision checks. In these papers, the standard setup involves one robot per octree node; however, in this study, two robots are permitted to occupy the same node. Collision checking is then exclusively performed between robots that are coexisting within a node or transitioning to a different node. Moreover, this arrangement offers an additional advantage by providing proximity information about robots that are close enough to be within the same quadrant but far enough not to collide within the next timestep. Leveraging this information, the quad-tree can be optimized by skipping collision checks and updates when applicable.

Regenerating quad-trees (RQ), which require the tree to be reconstructed at every timestep, can be computationally expensive. As a result, various methods have been explored to enable more efficient update mechanisms. \cite{Einhorn2011May} uses a generalization of quad-trees to generate a multi-resolution adaptive grid online based on the measurement uncertainty. This has shown to be effective in mapping, but the quad-tree update process is not conditioned on the states of the involved agents. \cite{Alamri2013Oct} used a tree structure for object tracking and assigned each object with an index to simplify the tree search process. The location of objects is only updated when the object moves into a new node. However, at each timestep, the new robot position still needs to be compared with the previous one to determine if the robot has entered a new node. In addition, several other papers have explored similar approaches for different applications, updating the tree only when necessary to track changes in the object's position \cite{Samet2013Jun} or at each timestep \cite{Saltenis2000Jun, Ding}. \cite{Kybic2010Aug} proposed a k-d tree algorithm to find the nearest neighbors in a set of points. Regions of the tree are updated when new points are added to the tree. Instead of updating the entire tree, the algorithm performs a local search around the newly added point to identify if the nearest neighbors to the other points have potentially changed. 

% \begin{figure}[htbp]
%     \centering 
%     \includegraphics[width=1\linewidth]{figures/quad_v4.png}
%     \caption{a). Only two robots are within the bounding box, which is equal to \textit{$m$}, so splitting the box is not required. b). A third robot is added to the bounding box and the total number of robots exceeds \textit{$m$}, so the space must be split into quadrants. c). An additional robot, C, is added. d). The space must again be split.}
%     \label{fig:quad_fig}
% \end{figure}

\section{Preliminaries}
\label{sec:preliminaries}
\subsection{System Definition}
The control objective is to regulate all agents from their initial states to their desired state in finite time. For our experiments, we use a dynamic bicycle model (refer to \cite{MPC_kinematic}), which is a good approximation for industrial wheeled robots. The model is discretized using Euler's method with a timestep of $\Delta t = 0.1$ s. We use a model predictive controller (MPC) similar to \cite{MPC_kinematic} as the motion planner for each agent. MPC only reasons about a single robot trajectory and does not perform collision avoidance, since this paper is focused on the collision detection problem. Note, our algorithm does not account for tracking or sensor errors.

A custom simulation written in Python was used for all the experiments and the collision checking algorithms. Experiments were done on a laptop with an AMD Ryzen 7 5700U with 22 GB of RAM. At every simulation step, each agent computes a sequence of control actions and predicted states using the MPC. Each agent then executes the next control action and updates its state according to the described dynamics model. Agents are assumed to have a circular footprint with a radius of $r = 0.5$.

\subsection{Quad-trees}

% \begin{figure}[tb]
%     \centering 
%     \includegraphics[width=1\linewidth]{figs/quad_v4.png}
%     \caption{quad-tree splitting examples. a) Only two robots are within the bounding box, which is equal to \textit{$m$}, so splitting the box is not required. b) A third robot is added to the bounding box and the total number of robots exceeds \textit{$m$}, so the space must be split into quadrants. c) An additional robot, C, is added. d) The space must again be split.}
%     \label{fig:quad_fig}
% \end{figure}

\begin{figure}[tb]
    \centering 
    \includegraphics[width=1\linewidth]{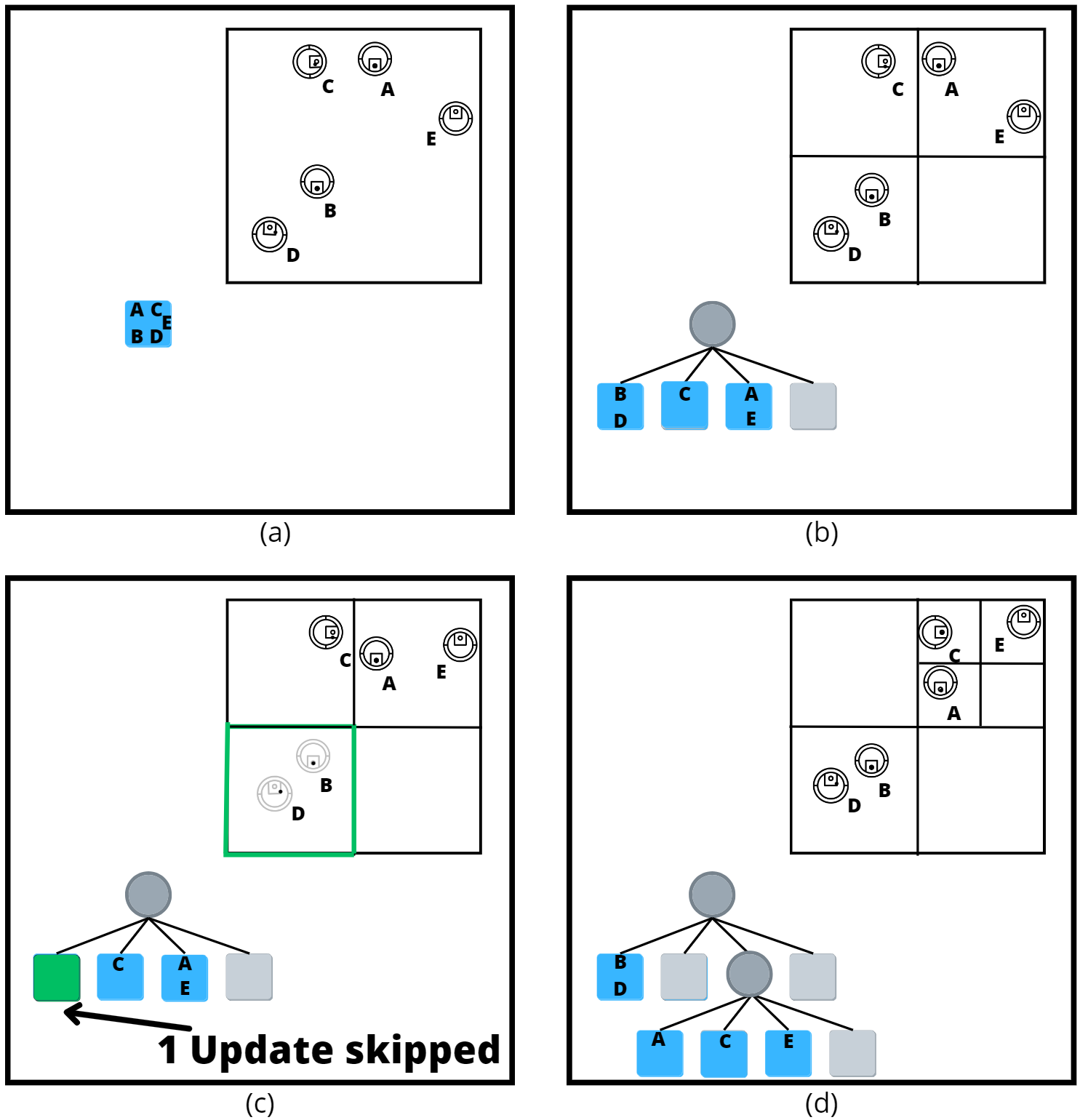}
    \caption{Example illustrating USQ in comparison to RQ. (a) Robots are placed in the first node of the regular quad-tree. (b) When the number of robots in the node exceeds the threshold $m$, where here $m=2$, the node is split into four quadrants. (c) In this step, USQ and RQ behave differently. In USQ, during the updating process, since robots B and D are far from the boundary and from each other updating the new robot positions into the quad-tree can be skipped. Updating the new positions of robots A, C, and E is not skipped because they are close to the boundary. In RQ, all the robot positions are updated in the quad-tree at each timestep (not shown in the figure for this step). (d) Since the robots are no longer far enough from each other to satisfy the update skipping condition (which is explained in detail in Section \ref{sec:USQ}), USQ behaves the same as RQ. All the robot positions are updated in the quad-tree.}
    \label{fig: USQ_Demo}
\end{figure}

The high-level idea of the quad-tree is to perform spatial partitioning of a given space to enable more efficient collision checking. This is done by successively dividing each partition into four equal-sized smaller partitions. These partitions can be described by a tree data structure, where each level of the tree going down from the root represents one level of division and each node a partition. This information can be used to restrict collision checking to only robots that are within close proximity. 

The algorithm starts by drawing a bounding box around all the robots. Let $m$ be the maximum number of robots allowed to be present within each partition. In this paper, this value is set to two robots. When there are more than $m$ robots occupying a partition, the quad-tree algorithm divides it into four new quadrants. This process is repeated until at most $m$ agents are in any given partition. Thus, each leaf node of the tree has at most $m$ agents at the end. An illustration of quad-trees can be seen in Fig. \ref{fig: USQ_Demo}. In this paper, a modified quad-tree data structure based on \cite{meijers_quadtree} is used.

\section{Updating and collision check Skipping Quad-tree Algorithm}
\label{sec:USQ}
\subsection{Overview}

% Algorithm 1 will be explained using the example of the quad-tree process in figure \ref{fig: USQ_Demo}. 

The USQ method is summarized in Algorithm 1 and Fig.~\ref{fig: USQ_Demo}. At the beginning of execution, an empty quad-tree is generated (Line 1). It is important to note that the full quad-tree is generated only once throughout the entire run and updated incrementally. This is in contrast to the RQ approach of regenerating the quad-tree at every timestep.

The following operations are then repeated until all the robots reach their final goal positions (Line 2). First, the MPC controller runs for each robot to find the next updated position of that robot (Line 4). Second, the algorithm must update the quad-tree structure (Lines 5-7) and then check for collisions (Lines 9-24).
At the end, the algorithm checks if all the robots have reached their goal positions (Line 25).

To optimize the efficiency of the algorithm, unnecessary updates and collision checks within a given quadrant are skipped. This is achieved by selectively excluding robots from updates and collision checks when they are unnecessary.
Each robot keeps track of a counter called $num\_skip$ which indicates the number of consecutive timesteps the robot can proceed without changing its position within the quad-tree or colliding with another robot. In essence, this counter determines how many quad-tree and collision check updates can be skipped in sequence. Skipping updates provides a significant computational saving over regenerating the quad-tree at every step, as each update of the tree requires traversing to a specific element. 
When the number of updates the robot can skip is 0 (Line 5), the robot position must be updated in the quad-tree (Line 6). 

The $update()$ step could be implemented a few different ways, but our implementation employs a remove-and-add strategy. This involves removing the robot from the quad-tree and subsequently placing it back in the updated location. Re-inserting the robot may require additional partitions to be added to the tree, which is done as needed.
Note that some of the tree nodes may become empty after removing robots and the original tree could be simplified by pruning those leaf nodes. However, we have found that the additional overhead required to perform this simplification outweighs the computational benefits. As a result, once a node is added to the tree it is kept in the structure throughout the execution. This effect can be seen in Fig.~\ref{fig:robot_swap}.

Once the quad-tree is updated with the latest locations of all robots, collision checking is performed. The algorithm iterates through all the leaf nodes of the quad-tree (Line 9) and checks for collisions between pairs of robots if any robot in that node has a $num\_skip=0$ (Line 10). In other words, collision checks are skipped when $num\_skip > 0$ for all robots in that node.
A collision between robots $a$ and $b$ with radii $r$ occurs when the following condition is true: 
\begin{equation}
collision(a,b):  \|a.pos - b.pos\| < 2r
\end{equation}
where $a.pos$, $b.pos$ are the robots' Cartesian positions. A more complicated collision checking algorithm, especially for non-circular robots, could be used at this line.

Next, if a robot could be close to a boundary of its quadrant,
we perform additional collision checks with neighboring nodes, since it may cross to a different quadrant at the subsequent timestep. To do so, we check if the distance to the quadrant's border $d\_border$ is below a specified minimum threshold $min\_thres$, (Line 15),
\begin{equation}
\label{eqn: minimum threshold}
    min\_thres = 2r + max\_dist\_traveled + \epsilon 
\end{equation}
$min\_thres$ takes into account the sum of the radii of two robots ($2r$) and the distance a robot can travel at maximum velocity in one timestep ($max\_dist\_traveled$). $\epsilon$ is a user-defined safety threshold. The equation represents the minimum distance a robot should maintain from the boundary to prevent potential collisions with a robot in the next quadrant during the next timestep. 
If the robot is within $min\_thres$ of the border, collision checks are done with the robots in the neighboring quadrant (Line 16). This process is explained in detail in Section \ref{subsec: Neighbor Checking}.

Next, the skipping counter $num\_skip$ must be updated. If $num\_skip$ is zero, then it is recalculated based on the minimum of two distances, $min\_d$: the distance of a robot to the border of its quadrant, $d\_border$, and the distance to other robots in its quadrant, $d\_robots$ (Line 18). Based on that minimum distance, $det\_skip()$ determines the number of quad-tree updates that can be skipped, if any, by computing the number of timesteps it would take the robot to collide with another robot or cross a boundary at maximum velocity (Line 19). If on the other hand $num\_skip$ is not yet zero, it is simply decremented by one (Line 21).

It is worth noting that the skipping of quad-tree updates and collision checks happen together in the USQ method. This is because one of the conditions for update skipping, robots being far from each other, aligns with the condition for collision check skipping.  Allowing the update and collision check skipping to occur separately incurred additional tracking costs that outweighed the benefits gained.

\begin{algorithm}[tb]
    \caption{USQ}
    \label{USQ Algo}
    \begin{algorithmic}[1]
       \State $T \gets quad\mhyphen tree(robot$\_$list$)
 \While{not $all\_reached\_goal$}

    \For{$robot$ $\in$ $robot$\_$list$}
        \State $robot.run$\_$mpc()$
        \If{$robot.num$\_$skip$ = 0}
            \State $T.update(robot)$
       
        \EndIf            
       
    \EndFor

    \For{$leaf \in T.nodes$}
        \If{$min(leaf.robots.num$\_$skip) = 0 $}
            \State $check$\_$collisions(leaf$\_$ node)$
        \EndIf
        \For{$robot \in leaf$}
        \If{$robot.num$\_$skip = 0$}
            \If{$d$\_$border  <  min$\_$thres$}
            \State $check$\_$neighbor$\_$collision(robot)$
        \EndIf
        \State $min$\_$d \gets min(d$\_$border, d$\_$robots)$
                \State $robot.num$\_$skip \gets det$\_$skip(min$\_$d)$    
               
        \Else
            \State $robot.num$\_$skip\gets robot.num$\_$skip- 1$
        \EndIf
        \EndFor
    \EndFor
   
    \State $all$\_$reached$\_$goal \gets check$\_$reached(robot$\_$list)$
\EndWhile
    \end{algorithmic}
\end{algorithm}

\subsection{Neighbor Checking}
\label{subsec: Neighbor Checking}
\begin{figure}[tb]
    \centering
    \subfigure[]{\includegraphics[width=0.1514\textwidth]{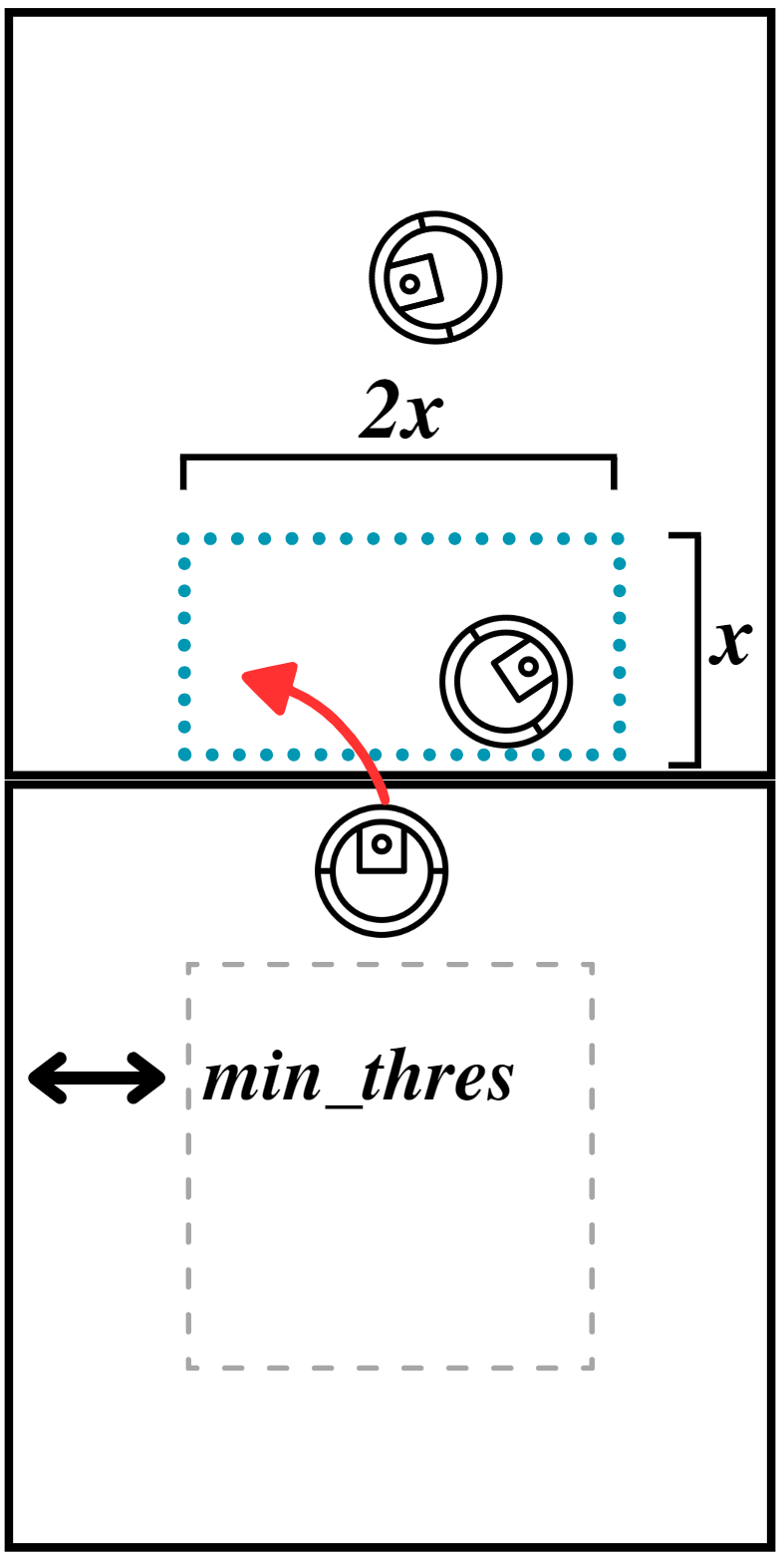}}
    \subfigure[]{\includegraphics[width=0.3\textwidth]{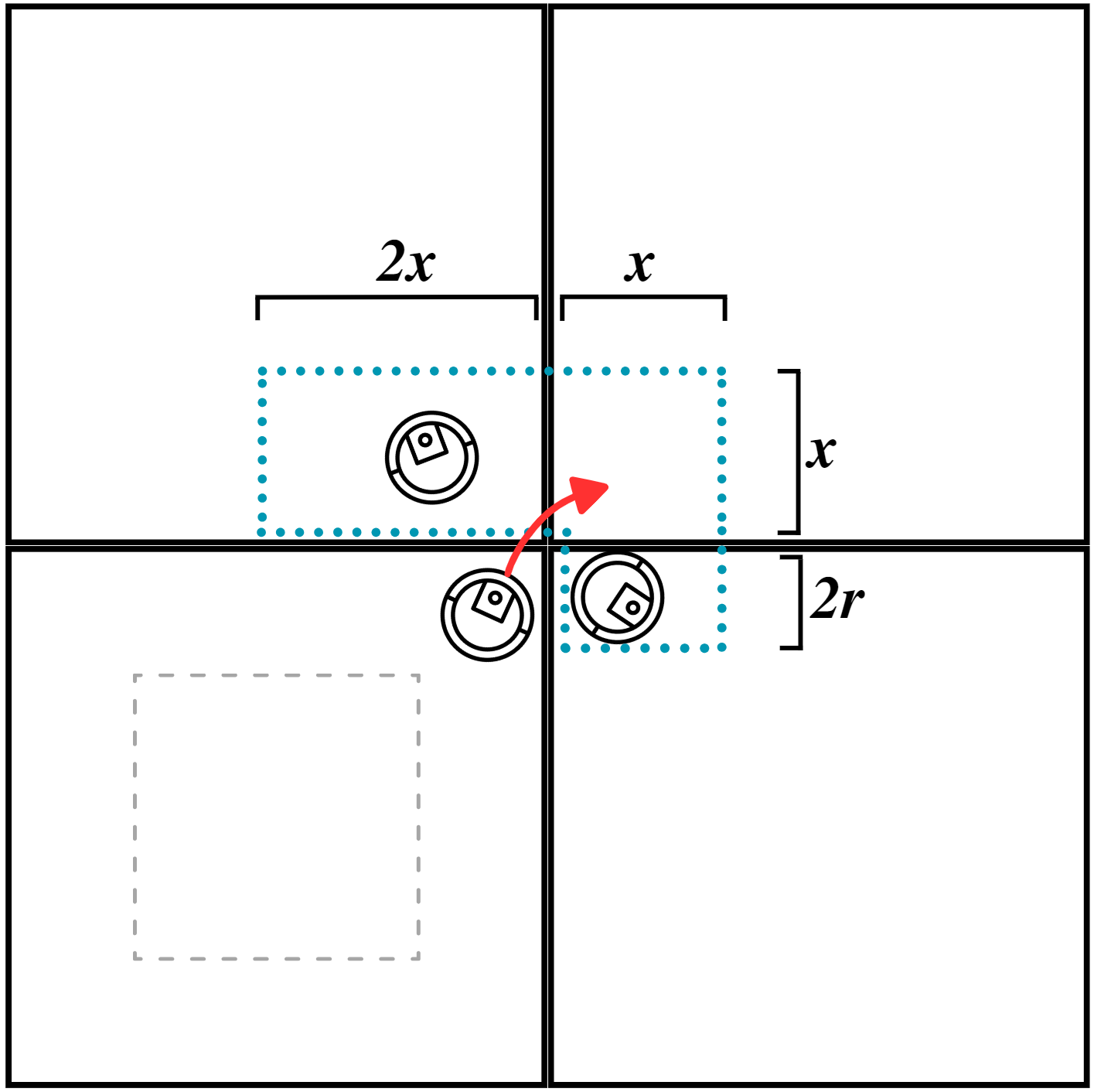}} 
    \caption{(a) When a robot approaches the node boundary, illustrated by its position beyond the gray dashed line, it does collision checks with robots situated within the adjacent node's blue dotted area. (b) When, in addition to the conditions described in (a), the robot's future trajectory intersects the quad-tree boundary at a linear distance of 2 radius units from the corner, it is identified as a corner case.}
    \label{fig: neighbor checking}
\end{figure}

Given that robots are positioned in the quad-tree based solely on their center position it is possible for them to collide at the boundary when their body extends beyond the edge or when transitioning to a neighboring quadrant. To address this issue, collision checks are performed on robots within a region of the upcoming quadrant when a robot is about to cross a boundary, as illustrated in Fig. \ref{fig: neighbor checking}(a).  This region is centered on the point that the robot's future trajectory intersects the quadrant border (or, if it will not cross within two timesteps, the region is centered on the closest point). It measures $2x$ by $x$, where $x$ is the distance in a given direction that a collision may occur. The region is $2x$ wide since the robot has the potential to travel a distance equal to $x$ to the left or right of the quadrant intersection point. Here we choose,
\begin{align}
    x= 2r + 2max\_dist\_traveled + \epsilon
\end{align}
since each robot can travel a distance equal to $max\_dist\_traveled$. 

Furthermore, if the robot's radius has the potential to extend into adjacent quadrants while crossing the boundary at a corner, collision checks are performed with robots present in two additional quadrants as shown in Fig. \ref{fig: neighbor checking}(b).  The bottom right search region has a height of $2r$ because, even though the robot's trajectory itself does not enter the region, its radius could potentially extend into that region while transitioning to another quadrant.

% Additionally, collision checks with robots in the three neighboring quadrants are conducted in cases that the robots' trajectory is expected to intersect any of the vertices.

\section{Results}
\label{sec:results}

\begin{figure}[t]
    \centering
    \subfigure[]{\includegraphics[width=0.225\textwidth]{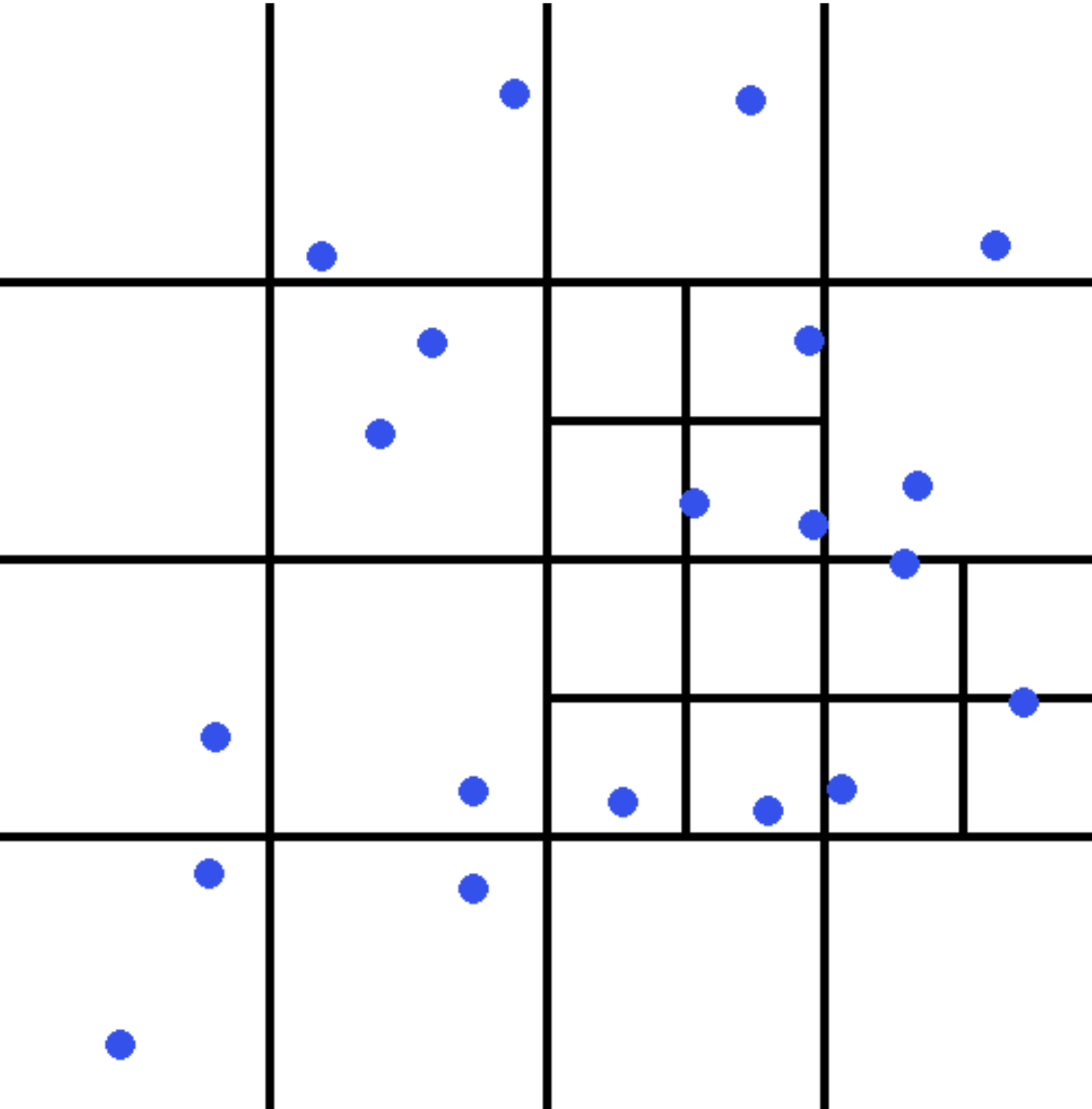}}
    \hspace{1em}
    \subfigure[]{\includegraphics[width=0.225 \textwidth]{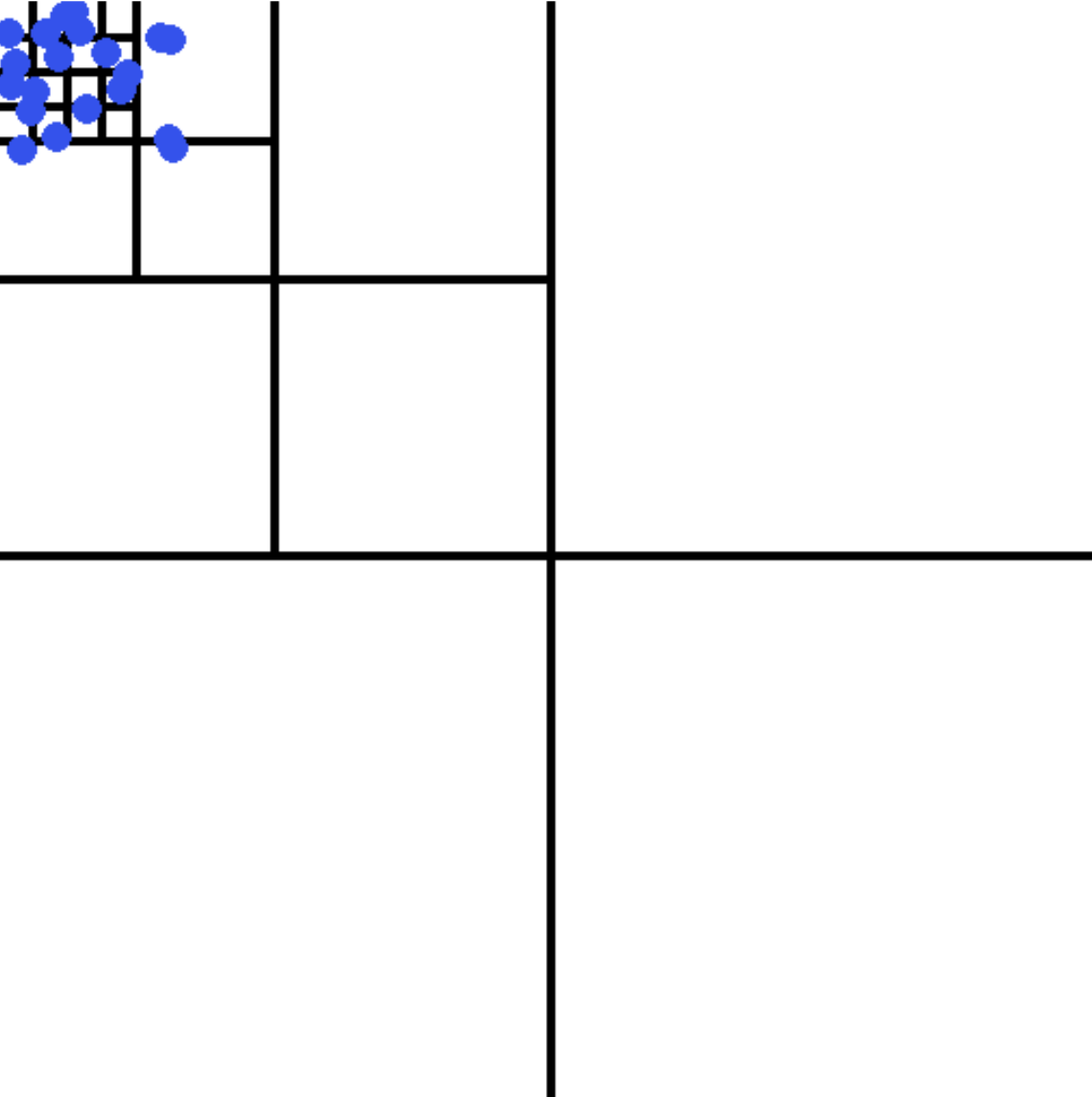}} 
    \caption{(a) Sparse and (b) dense environments with 20 robots.}
    \label{fig: env}
\end{figure}

In this section, we present results comparing the proposed USQ algorithm with RQ and pairwise collision checking across different numbers of robots and environments. RQ creates the quad-tree from scratch at each timestep and does not include any of the skipping logic, while the pairwise method checks for collisions between every pair of robots at each timestep regardless of their locations. 

We experiment with three separate environments: a sparse 512x512 unit environment (Fig. \ref{fig: env}(a)), a dense 85x85 unit environment (Fig. \ref{fig: env}(b)) and a circular 150 unit radius environment (Fig.~\ref{fig:robot_swap}). For each environment, we experiment with 5, 20, and 50 robots. The circle environment evaluates the performance over a particularly difficult single trial, whereas the sparse and dense environments are used to generate statistical results over 10 randomized trials for each test case. This results in a total of 60 random trials. For the randomized trials, the start and goal locations are randomly selected such that they are inside the map and non-overlapping. 

Multiple performance metrics were evaluated at the end of each run. $T_q$ is defined as the time spent creating and updating the quad-tree and $T_n$ is the time spent on neighbor checking, which includes traversing the tree to find neighboring robots and doing collision checks with them. $T_n$ provides insight into the additional computational cost of enabling quad-tree methods such as RQ and USQ achieve the same collision detection outcome as the more computationally expensive pairwise approach. The total collision checking time, $T_c$ encompasses all aspects associated with collision checking including the time spent on checking collisions between robots, quad-tree time ($T_q$), and neighbor checking time ($T_n)$, when appropriate. $N_c$ is the total number of collision checks, $N_d$ is the total number of collisions detected, and $N_m$ represents the number of collisions that were missed.

\subsection{Randomized Environments}

\begin{table}[t]
\setlength\tabcolsep{7pt}
\caption{Collision checking results in the sparse environment. Time is measured in seconds. Note: the dash (-) indicates values that are not applicable. For the $N_m$ column, it indicates the absence of collisions in the experiment and therefore, there are no collisions to be missed.}
\vspace{-1.5em}
\label{tab:sparse_env}
\begin{center}
\begin{tabular}{c|rrrrrr}
\multicolumn{7}{c}{} \\
\hline
 & \textbf{$T_q$} & \textbf{$T_n$} & \textbf{$T_c$} & \textbf{$N_c$} & \textbf{$N_d$} & \textbf{$N_m$} \\
\hline
\textbf{5 Robots} & & & & & & \\
Pairwise & \multicolumn{1}{c}{-} & \multicolumn{1}{c}{-} & \textbf{0.006} & 540 & 0 & \multicolumn{1}{c}{-} \\
RQ & 0.015 & \multicolumn{1}{c}{-} & 0.016 & 77 & 0 & \multicolumn{1}{c}{-} \\
USQ & 0.006 & 0.0001 & \textbf{0.006} & \textbf{5} & 0 & \multicolumn{1}{c}{-} \\
\hline
\textbf{20 Robots} & & & & & & \\
Pairwise & \multicolumn{1}{c}{-} & \multicolumn{1}{c}{-} & 0.083 & 10,260 & 0 & \multicolumn{1}{c}{-} \\
RQ & 0.063 & \multicolumn{1}{c}{-} & 0.065 & 264 & 0 & \multicolumn{1}{c}{-} \\
USQ & 0.040 & 0.0008 & \textbf{0.043} & \textbf{102} & 0 & \multicolumn{1}{c}{-} \\
\hline
\textbf{50 Robots} & & & & & & \\
Pairwise & \multicolumn{1}{c}{-} & \multicolumn{1}{c}{-} & 0.499 & 66,150 & \textbf{1} & \textbf{0} \\
RQ & 0.146 & \multicolumn{1}{c}{-} & 0.154 & 768 & \textbf{1} & \textbf{0} \\
USQ & 0.109 & 0.0072 & \textbf{0.121} & \textbf{445} & \textbf{1} & \textbf{0} \\
\end{tabular}
\end{center}
\end{table}

\begin{table}[t]
\setlength\tabcolsep{7pt}
\caption{Collision checking results in the dense environment. Time is measured in seconds. Note: the dash (-) indicates values that are not applicable. For the $N_m$ column, it indicates the absence of collisions in the experiment and therefore, there are no collisions to be missed.}
\vspace{-1.5em}
\label{tab:dense_env}
\begin{center}
\begin{tabular}{c|rrrrrr}
\multicolumn{7}{c}{} \\
\hline
 & \textbf{$T_q$} & \textbf{$T_n$} & \textbf{$T_c$} & \textbf{$N_c$} & \textbf{$N_d$} & \textbf{$N_m$} \\
\hline
\textbf{5 Robots} & & & & & & \\
Pairwise & \multicolumn{1}{c}{-} & \multicolumn{1}{c}{-} & \textbf{0.007} & 540 & 0 & \multicolumn{1}{c}{-} \\
RQ & 0.019 & \multicolumn{1}{c}{-} & 0.020 & 56 & 0 & \multicolumn{1}{c}{-} \\
USQ & 0.009 & 0.003 & 0.013 & \textbf{52} & 0 & \multicolumn{1}{c}{-} \\
\hline
\textbf{20 Robots} & & & & & & \\
Pairwise & \multicolumn{1}{c}{-} & \multicolumn{1}{c}{-}  & 0.084 & 10,260 & \textbf{8}& \textbf{0} \\
RQ & 0.067 & \multicolumn{1}{c}{-} & 0.071 & 323 & 6 & 2 \\
USQ & 0.047 & 0.014 & \textbf{0.063} & \textbf{228} & \textbf{8} & \textbf{0} \\
\hline
\textbf{50 Robots} & & & & & & \\
Pairwise & \multicolumn{1}{c}{-} & \multicolumn{1}{c}{-} & 0.505 & 66,150 & \textbf{56} & \textbf{0} \\
RQ & 0.148 & \multicolumn{1}{c}{-} & \textbf{0.160} & 1,090 & 42 & 14 \\
USQ & 0.128 & 0.055 & 0.194 & \textbf{928} & \textbf{56} & \textbf{0} \\
\end{tabular}
\end{center}
\end{table}

The results for the sparse environment, as shown in Table \ref{tab:sparse_env}, illustrate that USQ performs better than RQ for all cases in terms of the total number of collision checks ($N_c$) and total collision checking time ($T_c$).  As expected, pairwise collision checking does the worst for cases with many robots. However, for the 5-robot case, pairwise and USQ perform the same. This is because the additional computational overhead of the quad-tree does not outweigh the savings in collision checking time. As a result, we can conclude that pairwise collision checking is still an effective strategy for smaller robot teams. 

\begin{figure}[t]
    \centering 
    \includegraphics[width=0.65\linewidth]{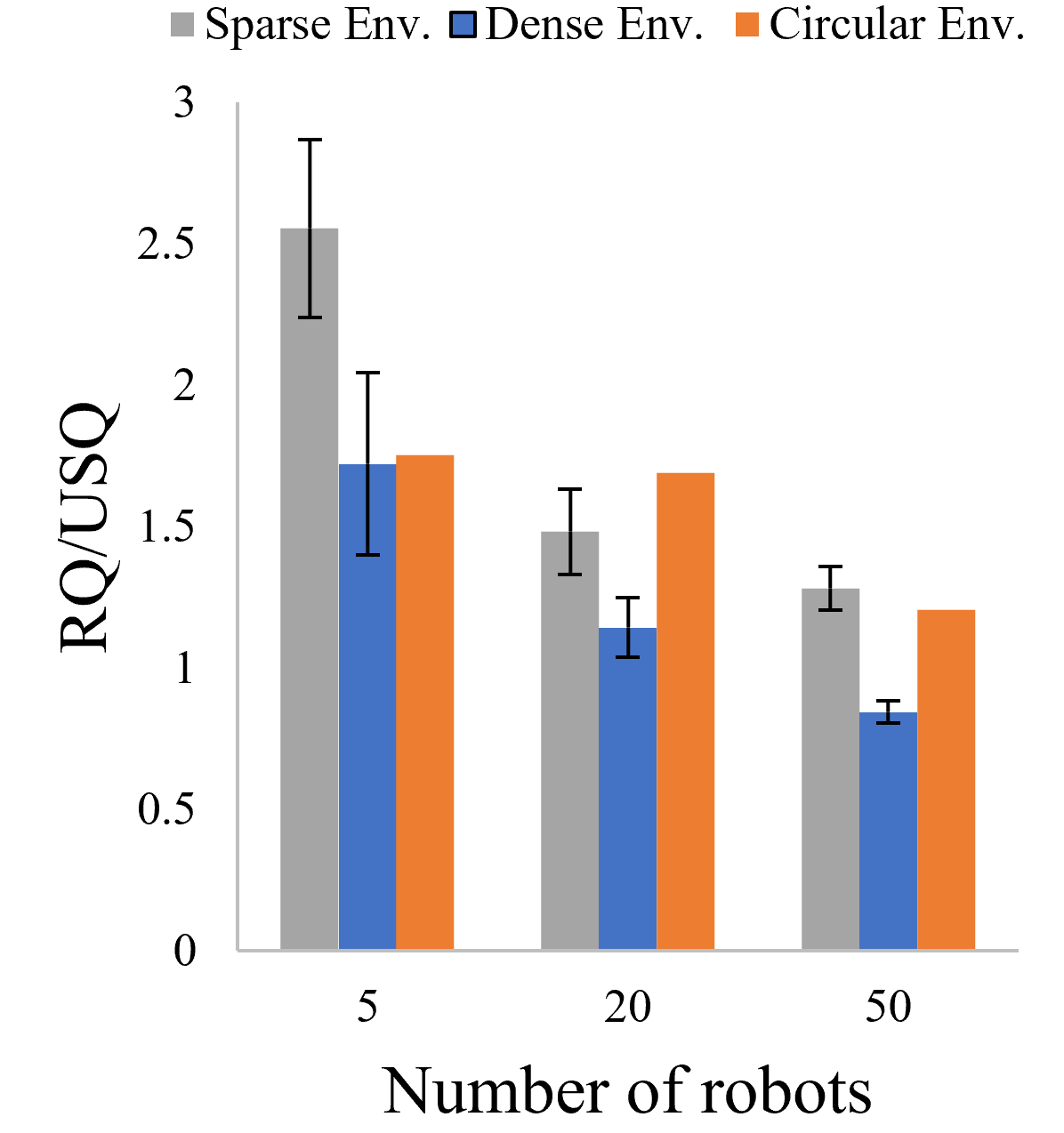}
    \caption{Speed-up time in the sparse, dense, and circular environments, defined as the ratio of $T_c$ values for RQ and USQ methods. Bars for the randomized test cases indicate $\pm$ one standard deviation.}
    \label{fig: speedup}
 \end{figure}

In the dense environment, as shown in Table \ref{tab:dense_env}, the trends are similar to the sparse environment. In the dense environment with 5 robots, pairwise actually outperforms both RQ and USQ. For larger numbers of robots, as demonstrated in Fig.~\ref{fig: speedup}, the speed-up in collision checking time between RQ and USQ is reduced compared to the sparse environment. Both of these trends are because, in the dense environment, agents are significantly closer to each other. Thus, fewer updates and collisions can be skipped. Additionally, the quad-tree quadrants are considerably smaller. This results in a higher number of agents crossing the boundaries and an increase in the number of collision checks with neighboring quadrants. For the 50-robot experiment, RQ has lower $T_c$ than USQ, although RQ fails to detect 25\% of the total collisions. While USQ still exhibits lower quad-tree update time ($T_q$) since some skipping occurs, the neighbor checking time ($T_n$) is relatively large. Despite this, neighbor checking is quite important to not miss any collisions.

Fig. \ref{fig: speedup} shows the speed-up of USQ when compared to RQ, showing the ratio of average $T_c$ for each. The speed-up is the smallest (0.82) in the denser environment with 50 robots, which has a higher robot density. Additionally, it is the greatest (2.6) in the sparse environment with 5 robots because the robots are more spread out, allowing more skips to occur. 

\subsection{Circular Environment}
\begin{figure}[tb]
    \centering 
    \includegraphics[width=.8\linewidth]{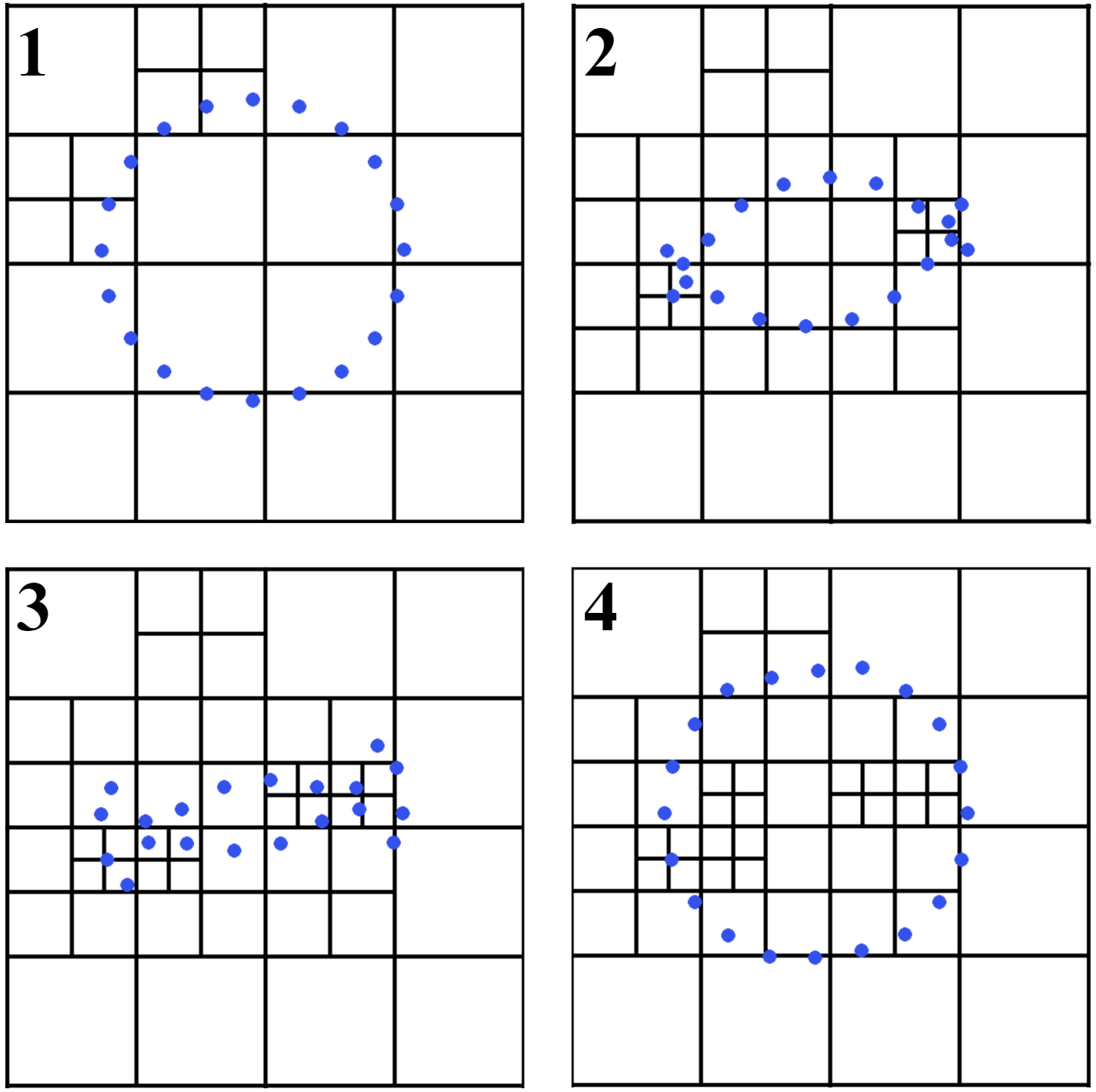}
    \caption{The circle environment, showing 20 robots swapping positions with a quad-tree implemented.}
    \label{fig:robot_swap}
\end{figure}

As shown in Fig. \ref{fig:robot_swap}, for this environment robots are placed in a circle with a radius of 150 units and are tasked to drive to a position mirrored across the $x$-axis from their starting position.
Table \ref{tab:circle} summarizes the results of this trial. Similar to the randomized environment cases, USQ and RQ perform worse than pairwise collision checking for the 5-robot case and USQ is better than both for the 20- and 50-robot cases. However, the advantage of USQ in the 50-robot case is smaller than in the 20-robot case since there are more robots close to each other which decreases the number of updates skipped. Fig. \ref{fig: speedup} shows the speed-up between RQ and USQ in the circle environment.

\begin{table}[t]
\setlength\tabcolsep{7pt}
\caption{Collision checking results in the circle environment. Time is measured in seconds. Note: the dash (-) indicates values that are not applicable.}
\vspace{-1.5em}
\label{tab:circle}
\begin{center}
\begin{tabular}{c|rrrrrr}
\multicolumn{7}{c}{} \\
\hline
 & \textbf{$T_q$} & \textbf{$T_n$} & \textbf{$T_c$} & \textbf{$N_c$} & \textbf{$N_d$} & \textbf{$N_m$} \\
\hline
\textbf{5 Robots} & & & & & & \\
Pairwise & \multicolumn{1}{c}{-} & \multicolumn{1}{c}{-} & \textbf{0.078} & 8,980 & \textbf{8} & \textbf{0} \\
RQ & 0.239 & \multicolumn{1}{c}{-} & 0.251 & 932 & \textbf{8} & \textbf{0} \\
USQ & 0.130 & 0.007 & 0.143 & \textbf{173} & \textbf{8} & \textbf{0} \\
\hline
\textbf{20 Robots} & & & & & & \\
Pairwise & \multicolumn{1}{c}{-} & \multicolumn{1}{c}{-} & 1.340 & 167,960 & \textbf{33} & \textbf{0} \\
RQ & 1.234 & \multicolumn{1}{c}{-} & 1.294 & 5,973 & 32& 1 \\
USQ & 0.682 & 0.037 & \textbf{0.766} & \textbf{3,284} &\textbf{33} & \textbf{0} \\
\hline
\textbf{50 Robots} & & & & & & \\
Pairwise & \multicolumn{1}{c}{-} & \multicolumn{1}{c}{-} & 8.511 & 1,097,600 & \textbf{87} & \textbf{0} \\
RQ & 2.857 & \multicolumn{1}{c}{-} & 3.007 & 15,085 & 81 & 6 \\
USQ & 2.071 & 0.325 & \textbf{2.494} & \textbf{9,742} & \textbf{87} & \textbf{0} \\
\end{tabular}
\end{center}
\end{table}

% \begin{figure}[!ht]
%     \centering 
%     \includegraphics[width=0.7\linewidth]{figs/circle speedup.png}
%     \caption{Speed-up time in the circle environment.}
%     \label{fig: circle speedup}
%  \end{figure}
     
% \begin{figure}[!ht]
%     \centering 
%     \includegraphics[width=0.75\linewidth]{figs/circle speedup.png}
%     \caption{Speed-up time in the circle environment}
%     \label{fig: circle speedup}
%  \end{figure}
 
\section{Conclusion}
\label{sec:conclusion}
This paper presents USQ, a novel and efficient collision checking algorithm for large multi-robot teams. Compared to RQ and pairwise collision checking, we demonstrate that USQ reduces collision checking time significantly in a variety of experiments with different numbers of robots and map sizes. USQ is found to be most effective in sparser scenarios, where robots form distinct clusters in large maps. However, there is still an improvement in denser scenarios compared to the other baselines. 

This research can be extended in multiple interesting directions. USQ can be used in multi-agent path finding and motion planning algorithms to speed up collision checking with other agents and obstacles. In addition, USQ can be extended to $3$-dimensional spaces to handle systems with higher degrees of freedom, such as drones.

%%%%%%%%%%%%%%%%%%%%%%%%%%%%%%%%%%%%%%%%%%%%%%%%%%%%%%%%%%%%%%%%%%%%%%%%%%%%%%%%

%%%%%%%%%%%%%%%%%%%%%%%%%%%%%%%%%%%%%%%%%%%%%%%%%%%%%%%%%%%%%%%%%%%%%%%%%%%%%%%%

\addtolength{\textheight}{-7.4cm}   % This command serves to balance the column lengths
\bibliographystyle{IEEEtran}
\bibliography{./IEEEfull,refs}

\begin{thebibliography}{10}
\providecommand{\url}[1]{#1}
\csname url@rmstyle\endcsname
\providecommand{\newblock}{\relax}
\providecommand{\bibinfo}[2]{#2}
\providecommand\BIBentrySTDinterwordspacing{\spaceskip=0pt\relax}
\providecommand\BIBentryALTinterwordstretchfactor{4}
\providecommand\BIBentryALTinterwordspacing{\spaceskip=\fontdimen2\font plus
\BIBentryALTinterwordstretchfactor\fontdimen3\font minus
  \fontdimen4\font\relax}
\providecommand\BIBforeignlanguage[2]{{%
\expandafter\ifx\csname l@#1\endcsname\relax
\typeout{** WARNING: IEEEtran.bst: No hyphenation pattern has been}%
\typeout{** loaded for the language `#1'. Using the pattern for}%
\typeout{** the default language instead.}%
\else
\language=\csname l@#1\endcsname
\fi
#2}}

\bibitem{sharon2012}
G.~Sharon, R.~Stern, A.~Felner, and N.~Sturtevant, ``Meta-agent conflict-based
  search for optimal multi-agent path finding,'' in \emph{International
  Symposium on Combinatorial Search}, vol.~3, no.~1, 2012.

\bibitem{pairwise_big_o_reference}
M.~H.~J. Saldanha and P.~S. L.~d. Souza, ``High performance algorithms for
  counting collisions and pairwise interactions,'' in \emph{International
  Conference on Computational Science}.\hskip 1em plus 0.5em minus 0.4em\relax
  Springer, 2019, pp. 182--196.

\bibitem{gao2017gradient}
F.~Gao, Y.~Lin, and S.~Shen, ``Gradient-based online safe trajectory generation
  for quadrotor flight in complex environments,'' in \emph{IEEE/RSJ
  International Conference on Intelligent Robots and Systems}, 2017, pp.
  3681--3688.

\bibitem{chen2016online}
J.~Chen, T.~Liu, and S.~Shen, ``Online generation of collision-free
  trajectories for quadrotor flight in unknown cluttered environments,'' in
  \emph{IEEE International Conference on Robotics and Automation}, 2016, pp.
  1476--1483.

\bibitem{tordesillas2021faster}
J.~Tordesillas, B.~T. Lopez, M.~Everett, and J.~P. How, ``Faster: Fast and safe
  trajectory planner for navigation in unknown environments,'' \emph{IEEE
  Transactions on Robotics}, vol.~38, no.~2, pp. 922--938, 2021.

\bibitem{Botea2004Jan}
A.~Botea, M.~M{\ifmmode\ddot{u}\else\"{u}\fi}ller, and J.~Schaeffer, ``Near
  optimal hierarchical path-finding {(HPA{$\ast$})},'' \emph{Journal of Game
  Development}, vol.~1, Jan. 2004.

\bibitem{Kambhampati1986Sep}
S.~Kambhampati and L.~Davis, ``Multiresolution path planning for mobile
  robots,'' \emph{IEEE Journal on Robotics and Automation}, vol.~2, no.~3, pp.
  135--145, Sept. 1986.

\bibitem{Kitamura1995Aug}
Y.~Kitamura, T.~Tanaka, F.~Kishino, and M.~Yachida, ``{3-D} path planning in a
  dynamic environment using an octree and an artificial potential field,'' in
  \emph{IEEE/RSJ International Conference on Intelligent Robots and Systems},
  Aug. 1995, vol.~2, pp. 474--481vol.2.

\bibitem{Ding}
R.~Ding and X.~Meng, ``A quadtree based dynamic attribute index structure and
  query process,'' in \emph{International Conference on Computer Networks and
  Mobile Computing}.\hskip 1em plus 0.5em minus 0.4em\relax IEEE, 2001, pp.
  16--19.

\bibitem{Alamri2013Oct}
S.~Alamri, D.~Taniar, M.~Safarb, and H.~Al-Khalidi, ``Tracking moving objects
  using topographical indexing,'' \emph{Concurrency and Computation: Practice
  and Experience}, vol.~27, Oct. 2013.

\bibitem{Honig2018Aug}
W.~H{\ifmmode\ddot{o}\else\"{o}\fi}nig, J.~A. Preiss, T.~K.~S. Kumar,
  \emph{et~al.}, ``Trajectory planning for quadrotor swarms,'' \emph{IEEE
  Transactions on Robotics}, vol.~34, no.~4, pp. 856--869, Aug. 2018.

\bibitem{Wu2022May}
X.~Wu, Y.~Lei, X.~Tong, \emph{et~al.}, ``A non-rigid hierarchical discrete grid
  structure and its application to {UAVs} conflict detection and path
  planning,'' \emph{IEEE Transactions on Aerospace and Electronic Systems},
  vol.~58, no.~6, pp. 5393--5411, May 2022.

\bibitem{Dames2012Dec}
P.~Dames, M.~Schwager, V.~Kumar, and D.~Rus, ``A decentralized control policy
  for adaptive information gathering in hazardous environments,'' in \emph{IEEE
  Conference on Decision and Control}, Dec. 2012, pp. 2807--2813.

\bibitem{Collins2021May}
L.~Collins, P.~Ghassemi, E.~T. Esfahani, \emph{et~al.}, ``Scalable coverage
  path planning of multi-robot teams for monitoring non-convex areas,'' in
  \emph{IEEE International Conference on Robotics and Automation}, May 2021,
  pp. 7393--7399.

\bibitem{Johansen2003May}
T.~A. Johansen and A.~Grancharova, ``Approximate explicit constrained linear
  model predictive control via orthogonal search tree,'' \emph{IEEE
  Transactions on Automatic Control}, vol.~48, no.~5, pp. 810--815, May 2003.

\bibitem{Cychowski2005Aug}
M.~T. Cychowski and T.~O'Mahony, ``Efficient approximate robust {MPC} based on
  quad-tree partitioning,'' in \emph{IEEE Conference on Control Applications},
  Aug. 2005, p. 239.

\bibitem{yijun2021fast}
Z.~Yijun, X.~Jiadong, and L.~Chen, ``A fast bi-directional {A*} algorithm based
  on quad-tree decomposition and hierarchical map,'' \emph{IEEE Access},
  vol.~9, pp. 102\,877--102\,885, 2021.

\bibitem{Singh2017Oct}
P.~Singh and S.~K. Vishvakarma, ``Low complexity-low power object tracking
  using dynamic quadtree pixelation and macroblock resizing,'' \emph{Pattern
  Recognition and Image Analysis}, vol.~27, no.~4, pp. 731--739, Oct. 2017.

\bibitem{Datcu1992Nov}
M.~P. Datcu, F.~Folta, and C.~E. Toma, ``Algorithm for dynamic object
  tracking,'' in \emph{Intelligent Robots and Computer Vision XI: Algorithms,
  Techniques, and Active Vision}.\hskip 1em plus 0.5em minus 0.4em\relax SPIE,
  Nov. 1992, vol. 1825, pp. 389--394.

\bibitem{log(n)}
R.~A. Finkel and J.~L. Bentley, ``Quad trees a data structure for retrieval on
  composite keys,'' \emph{Acta informatica}, vol.~4, no.~1, pp. 1--9, 1974.

\bibitem{Rodenberg2016Sep}
O.~Rodenberg, E.~Verbree, and S.~Zlatanova, ``Indoor {A$\ast$} pathfinding
  through an octree representation of a point cloud,'' \emph{ISPRS Annals of
  the Photogrammetry, Remote Sensing and Spatial Information Sciences}, vol.
  IV-2/W1, Sept. 2016.

\bibitem{Zhang2022}
T.~Zhang, C.~Li, X.~J. Zhu, \emph{et~al.}, ``A multi-robot planning algorithm
  with quad tree map division for obstacles of irregular shape,'' in
  \emph{Design Studies and Intelligence Engineering}.\hskip 1em plus 0.5em
  minus 0.4em\relax IOS Press, 2022, pp. 24--32.

\bibitem{csenbacslar2019robust}
B.~{\c{S}}enba{\c{s}}lar, W.~H{\"o}nig, and N.~Ayanian, ``Robust trajectory
  execution for multi-robot teams using distributed real-time replanning,'' in
  \emph{Distributed Autonomous Robotic Systems: The 14th International
  Symposium}.\hskip 1em plus 0.5em minus 0.4em\relax Springer, 2019, pp.
  167--181.

\bibitem{Einhorn2011May}
E.~Einhorn, C.~Schr{\ifmmode\ddot{o}\else\"{o}\fi}ter, and H.-M. Gross,
  ``Finding the adequate resolution for grid mapping - {Cell} sizes locally
  adapting on-the-fly,'' in \emph{IEEE International Conference on Robotics and
  Automation}, May 2011, pp. 1843--1848.

\bibitem{Samet2013Jun}
H.~Samet, J.~Sankaranarayanan, and M.~Auerbach, ``Indexing methods for moving
  object databases: Games and other applications,'' \emph{ACM SIGMOD
  International Conference on Management of Data}, pp. 169--180, June 2013.

\bibitem{Saltenis2000Jun}
S.~{\ifmmode\check{S}\else\v{S}\fi}altenis, C.~Jensen, S.~Leutenegger, and
  M.~Lopez, ``Indexing the positions of continuously moving objects,''
  \emph{Sigmod Record}, vol.~29, pp. 331--342, June 2000.

\bibitem{Kybic2010Aug}
J.~Kybic and I.~Vnu{\ifmmode\check{c}\else\v{c}\fi}ko, ``Approximate best bin
  first k-d tree all nearest neighbor search with incremental updates,'' Czech
  Technical University in Prague, Tech. Rep. K333-40/10, CTU-CMP-2010-10, 2010.

\bibitem{MPC_kinematic}
J.~Kong, M.~Pfeiffer, G.~Schildbach, and F.~Borrelli, ``Kinematic and dynamic
  vehicle models for autonomous driving control design,'' in \emph{IEEE
  Intelligent Vehicles Symposium}, 2015, pp. 1094--1099.

\bibitem{meijers_quadtree}
B.~Meijers, ``Quadtree,'' \url{https://github.com/bmmeijers/quadtree}, 2021,
  [Accessed: Dec. 20, 2023].

\end{thebibliography}

\end{document}